\documentclass[letter]{aa}

\usepackage{txfonts}
\usepackage{graphicx}
\usepackage{natbib}

\titlerunning{Instabilities in relativistic blast waves}
\authorrunning{Z. Meliani et al.}

\begin{document}
\title{Dynamics and stability of relativistic GRB blast waves} 

  \author{Z. Meliani\inst{1}, R. Keppens \inst{1,2,3}}
\offprints{Z. Meliani}
   \institute{Centre for Plasma Astrophysics, Celestijnenlaan 200B, 3001 Heverlee, K.U.Leuven, Belgium \and 
FOM-Institute for Plasma Physics Rijnhuizen, Nieuwegein, The Netherlands \and
Astronomical Institute, Utrecht University, The Netherlands\\
              \email{zakaria.meliani@wis.kuleuven.be, rony.keppens@wis.kuleuven.be}}

   \date{Received ; accepted }

\abstract{}
{In gamma-ray-bursts (GRB), ultra-relativistic blast waves are ejected into the circumburst medium. We analyse in unprecedented detail the deceleration of a self-similar Blandford-McKee blast wave from a Lorentz factor 25 to the nonrelativistic Sedov phase. Our goal is to determine the stability properties of its frontal shock.}{We carried out a grid-adaptive relativistic 2D hydro-simulation at extreme resolving power, following the GRB jet during the entire afterglow phase. We investigate the effect of the finite initial jet opening angle on the deceleration of the blast wave, and identify the growth of various instabilities throughout the coasting shock front.}{We find that during the relativistic phase, the blast wave is subject to pressure-ram pressure instabilities that ripple and fragment the frontal shock. These instabilities manifest themselves in the ultra-relativistic phase alone, remain in full agreement with causality arguments, and decay slowly to finally disappear in the near-Newtonian phase as the shell Lorentz factor drops below 3. From then on, the compression rate decreases to levels predicted to be stable by a linear analysis of the Sedov phase.   
Our simulations confirm previous findings that the shell also spreads laterally because a rarefaction wave slowly propagates to the jet axis, inducing a clear shell deformation from its initial spherical shape. The blast front becomes meridionally stratified, with decreasing speed from axis to jet edge.
In the wings of the jetted flow, Kelvin-Helmholtz instabilities occur, which are of negligible importance from the energetic viewpoint.}{Relativistic blast waves are subject to hydrodynamical instabilities that can significantly affect their deceleration properties. Future work will quantify their effect on the afterglow light curves.
}

\maketitle
\section{Progress in GRB dynamics modeling}

Gamma ray burst (GRB) outflows represent the most extreme variant of relativistic jets. That they represent finite opening angle flows is supported by the appearance of jet breaks in GRB afterglows~\citep{Frailetal97} (insights provided by a coupled simulation and light-curve synthesis point out subtle chromatic effects, van Eerten et al. 2010). Other convincing arguments rely on the polarization in prompt emission \citep{Lazzatietal04}, as well as theoretical reasoning to invoke physically plausible explosion energies \citep{Bloometal03}.

However, despite the progress in GRB understanding allowed by current detailed observations, various questions remain. In particular, for the (micro and macro) physics occurring within the relativistic shock fronts. To improve constraints on GRB progenitor stars and overall GRB energies, we need to develop improved modeling, targeting the interaction of collimated relativistic blast waves with circumburst progenitor surroundings. This interaction is subject to non-linear physics affecting the details of how energy is transferred from the collimated ejecta to the circumburst medium. Studies of this interaction rely on accurate, shock-capturing high resolution simulations. Previous 1D studies using Lorentz factor 100 shells in either uniform~\citep{Melianietal07} or fully wind-structured surroundings~\citep{zaklet07} quantified the need to resolve a scale ratio of up to $6$ orders of magnitude between shell width and traversed distance, to study the blastwave until the Sedov phases.

Pioneering work in GRB dynamics in afterglow phases was perfomed analytically, e.g. \cite{Meszaros&Rees97}, and numerically, e.g. \cite{Kobayashietal99}, using isotropic explosion models. \cite{Rhoads99} studied GRB outflow in two-dimensional settings analytically, by making assumptions anout sideways jet-expansion speeds. In the past decade, two and even three-dimensional  relativistic simulations emerged, necessarily sacrificing in resolution \citep{Granotetal01, cannizzoetal04}, grid-adaptive studies at very high effective resolution \citep{ZhangMacFadyen09,eerten10} as well as high initial Lorentz factors (up to $\gamma=100$ Meliani et al. 2007) became feasible. In contrast to earlier analytic assumptions, these simulations found a slow lateral expansion of the GRB jet, in accord with semi-analytical works \citep{Kumar&Granot03}. However, the stability properties of the decelerating relativistic blast wave have not yet been addressed. in contrast, to the Newtonian regime, where the self-similar Sedov blast-wave structure is known to be unstable~\citep{Vishniac83} when the compression rate is high (in turn a function of polytropic index $\Gamma<1.3$), the relativistic regime is more complex since the compression rate depends on effective polytropic index, as well as on the Lorentz factor. Relativistic blast-wave stability has to be studied in a single dimension, and can only be performed at extreme resolutions, as demonstrated here.

\begin{figure*}
\begin{center}
{\resizebox{1.8\columnwidth}{9.0cm}{\includegraphics{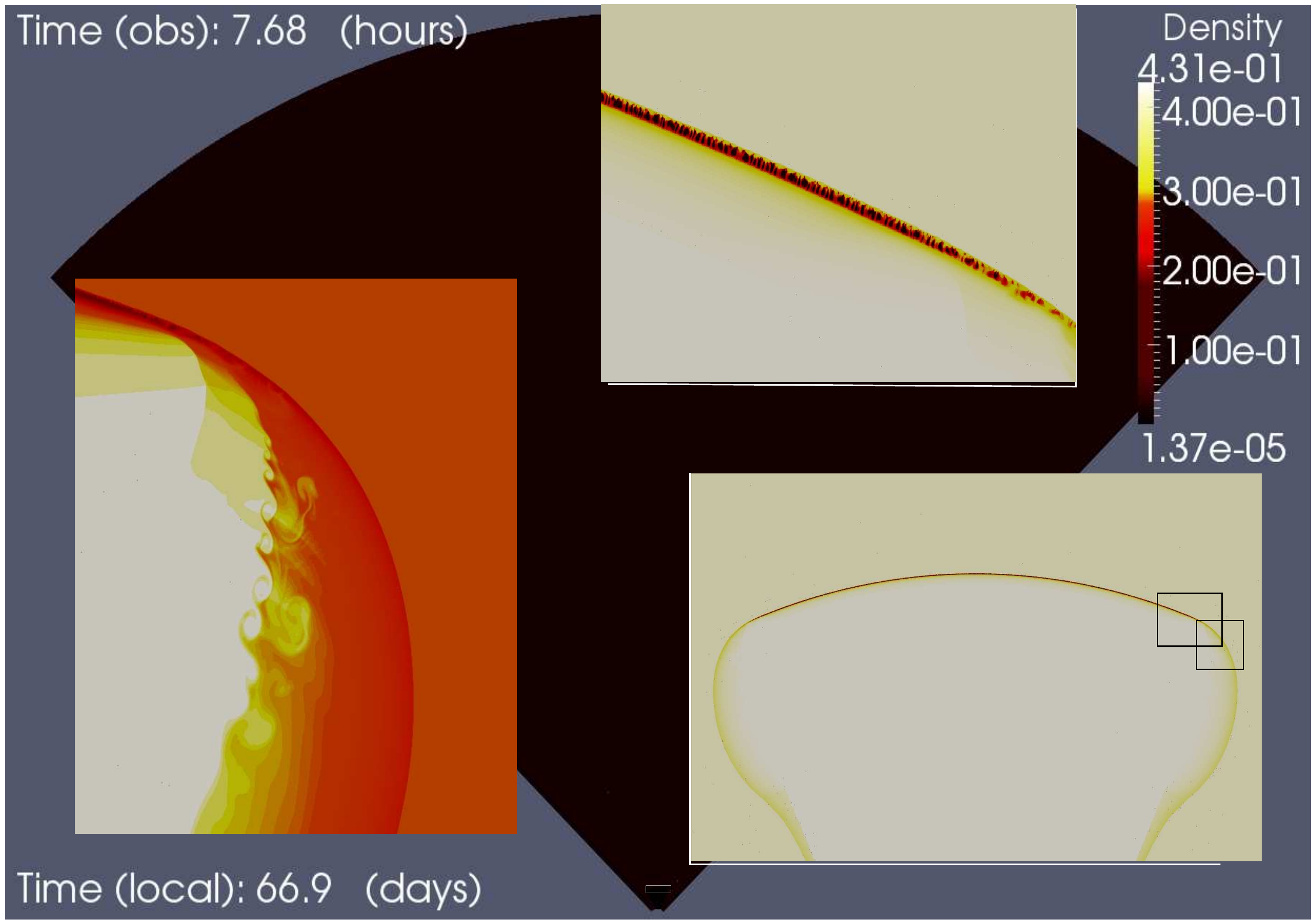}}}
\end{center}
\vspace{-0.2cm}
\caption{The density variation at local time $t=66.9 \,{\rm days}$, with the large-scale view in the background, and a zoom into the shell at the bottom right (zooming into the square at the bottom of the large-scale view), yet further into zoomed to the shell edge at top right and in bottom left (zoomed on the squares in the full shell view). In the zoom figures we use inverse colors.}\label{Figall}
\end{figure*}

\section{Model setup}\label{modelone}
\vspace*{-0.25cm}

The two-dimensional hydrodynamics blast wave simulation in this paper was performed with the special relativistic hydro-module of the adaptive-mesh refinement code AMRVAC~\citep{Melianietal07}. We use a Synge-type equation of state \citep{Melianietal04} that takes full account of the change in effective polytropic index.  In the present paper we neglect the effect of magnetic fields and radiative losses.
As an initial condition, a  self-similar relativistic~\cite{BlandfordMcKee76} blast wave structure, given by their equations Eqs~(28)-(30), is assumed to have a half-opening angle $\theta=20^{\circ}$,
 energy $E_{\rm jet}\,=\,E_{\rm iso} \theta^2/2 \,{\rm ergs}$, and equivalent isotropic energy  $E_{\rm iso}\,=\,2.6\times10^{51} \,{\rm ergs}$. 
The blast wave propagates in a constant density medium $n\,=\,0.78 {\rm cm^{-3}}$. 
We begin the simulation at a shock Lorentz factor $\gamma_{sh}\,=\,25$ (implying a local Lorentz factor of $\gamma\simeq 25/\sqrt{2}\simeq 17.7$), making the initial radius of the blast wave $R_{0}\,\simeq\, 2.68\, \times\, 10^{-2}\,{\rm pc}$ at local time $t_{\rm local,0}\,\simeq\, 51.68\, {\rm days}$ and observer time $t_{\rm obs,0}\,=R_0/2\gamma^2 c \simeq 1 \,{\rm hour}$. At that time, the blast wave energy is mostly concentrated in a thin layer $\Delta R\,\simeq \,R_0/\gamma^2=8.5\times 10^{-5} {\rm pc}$. To follow the evolution of this blast wave from relativistic to Newtonian phases, we use a spherical domain spanning $\left[1.57\times 10^{-2},\,1.57\right]\;{\rm pc}$ in radius and $\left[0,45^{\circ}\right]$ in meridional angle. To simulate this blast wave over this large dynamical range, we need adaptive mesh refinement (AMR). This AMR run exploits 13 grid levels  and uses an error estimator for normalized second derivatives on density with a weight of $0.9$ and of pressure with a weight of $0.1$ as criterion for refinement/coarsening. In addition, we enforce coarsening in the back of the shell, and begin with $280 \times 28$ grid cells at level 1. By using a refinement ratio 2 between levels, we achieve an effective resolution of $\left(1.14688 \times 10^{6}\right) \times  \left(1.14688 \times 10^{5}\right)$.  Our block-tree AMR automatically tracks the shock front, while the number of blocks remains nearly constant during the run: we have typically  8 blocks at the lowest level and about 50 000 blocks at the highest level. With this extreme resolving power, we can capture radial details at $\delta R \simeq 1.37 \times 10^{-6}  {\rm pc}$.
In the meridional direction, we have $\delta \theta \simeq 0.0004^{\circ}$, and the initial shell contains $50 972$ meridional cells. Figure~\ref{Figall} illustrates all the scales that enter this simulation.  We used a predictor-corrector scheme with the HLLE solver~\citep{Hartenetal83} and minmod limited linear reconstruction. This scheme avoids potential numerical instabilities at strong shocks~\citep{Mulleretal10}, as confirmed by analysis of its linearized form~\citep{Pandolfi&DAmbrosio01}.

\begin{figure}
\begin{center}
{\resizebox{0.95\columnwidth}{4.8cm}{\includegraphics{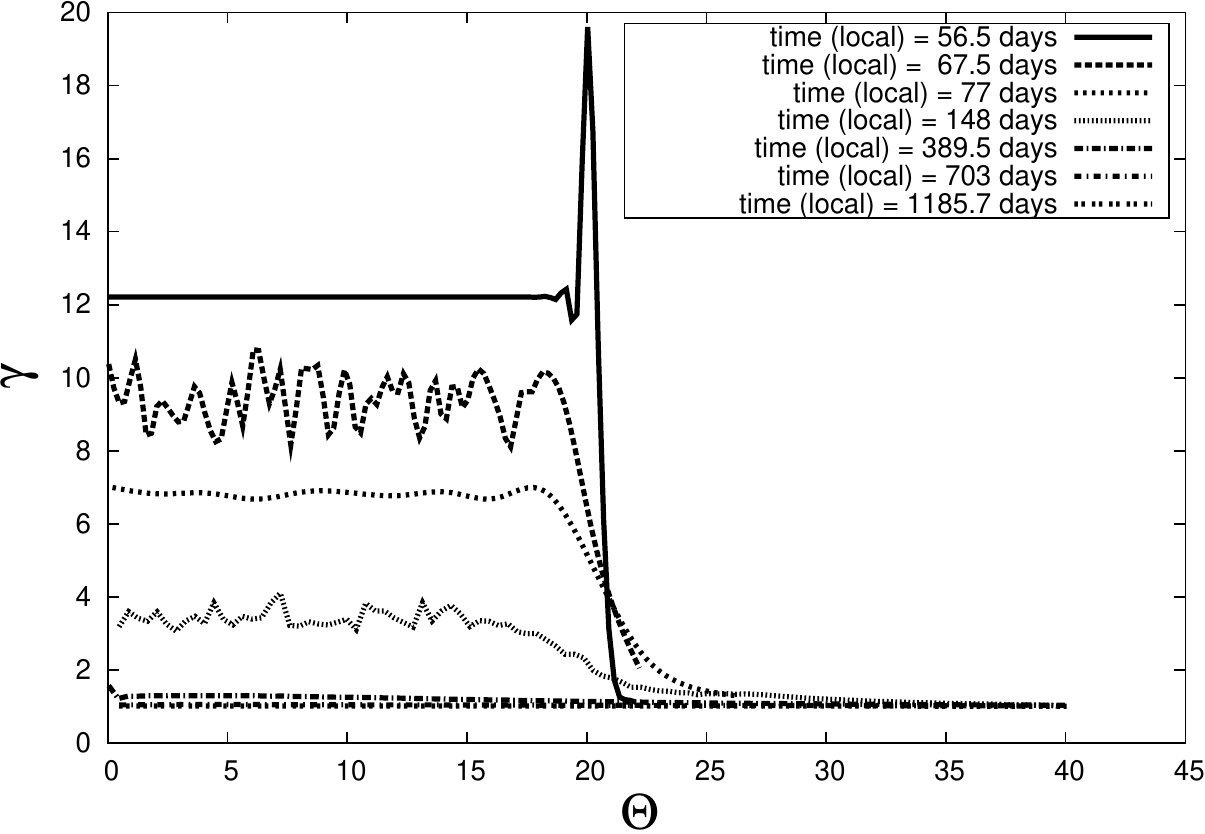}}}
\end{center}
\vspace{-0.3cm}
\caption{The variation in the Lorentz factor at the front shock against angle $\theta$, at various times in the blast wave evolution.}\label{Figlfac}
\vspace{-0.4cm}
\end{figure}
\begin{figure}[h]
\begin{center}
{\resizebox{0.95\columnwidth}{4.8cm}{\includegraphics{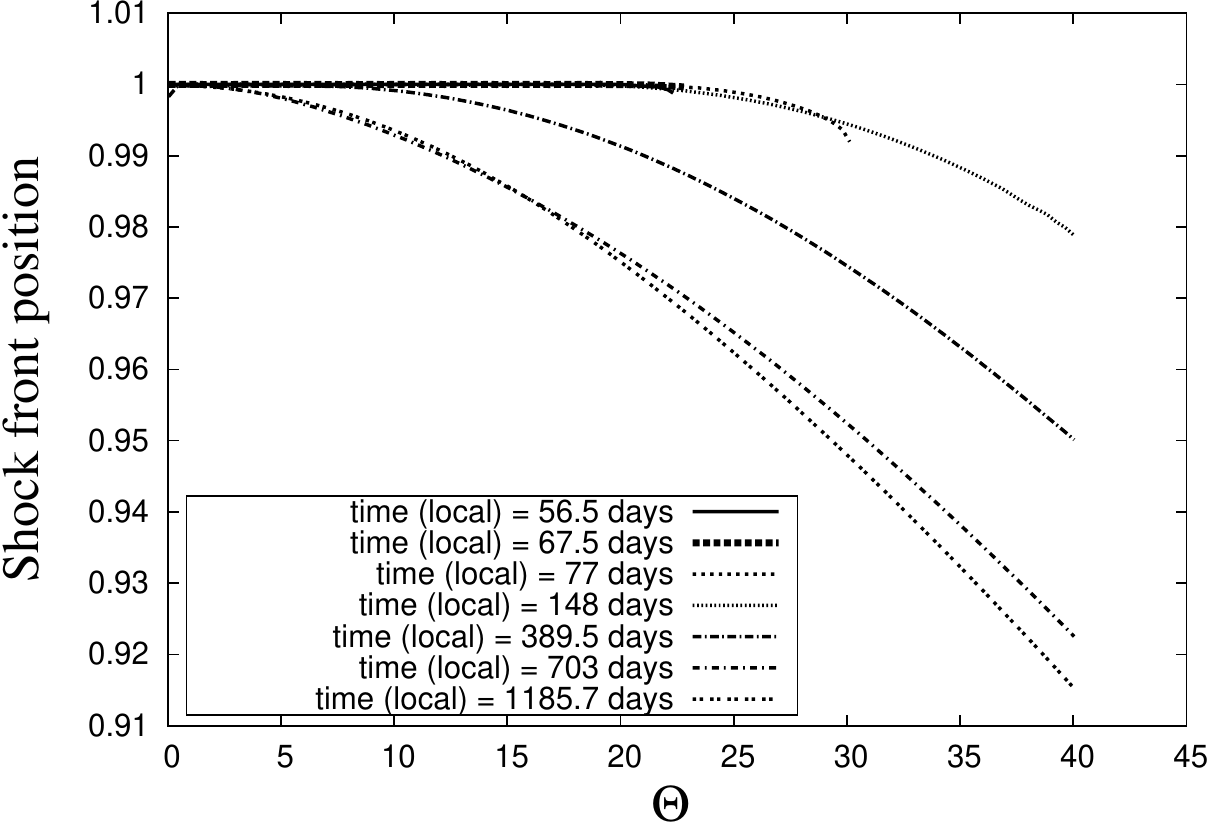}}}
\end{center}
\vspace{-0.5cm}
\caption{The variation in the radial position of the frontal shock, normalized to show a maximum unit value at each time, as a function of angle $\theta$ from axis. This shows the front deformation with time.}\label{Blast_R}
\vspace{-0.4cm}
\end{figure}
\begin{figure}
\begin{center}
{\resizebox{0.95\columnwidth}{4.8cm}{\includegraphics{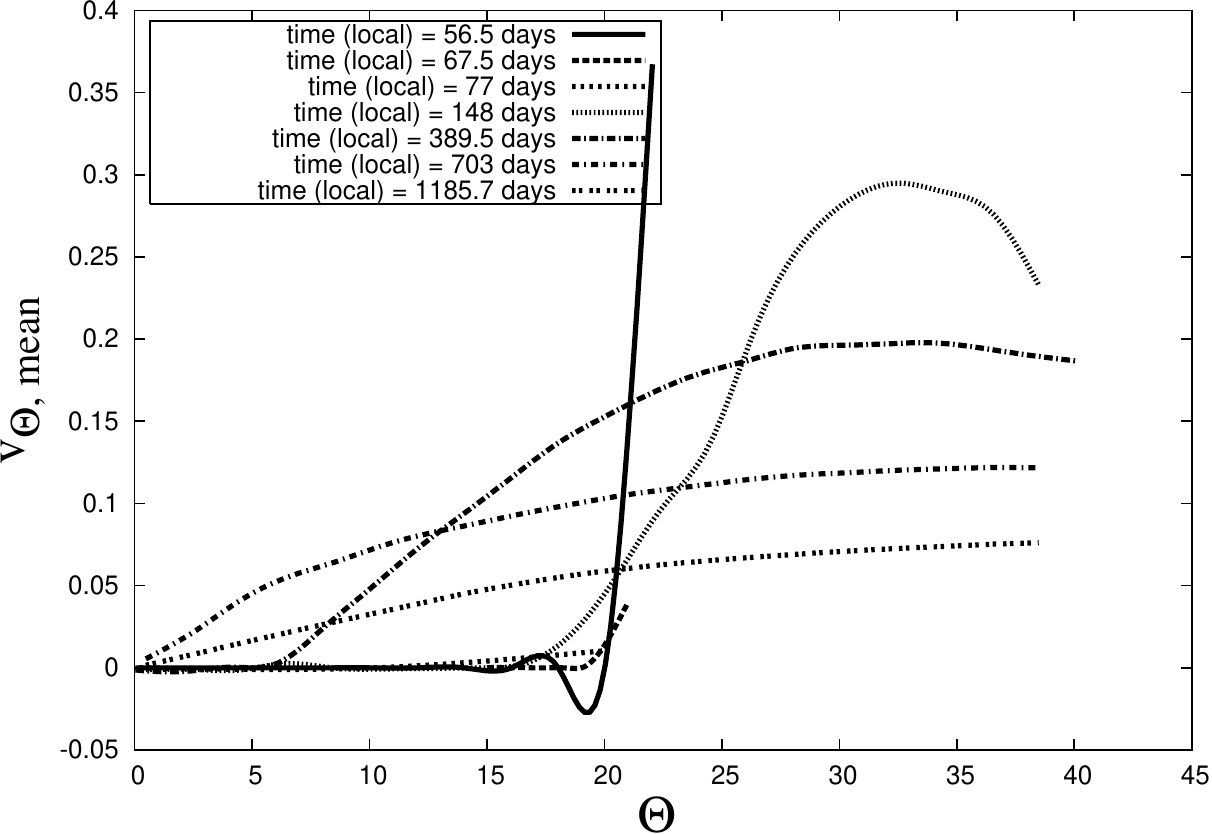}}}
\end{center}
\vspace{-0.7cm}
\caption{The variation in meridional speed at the shock front with angle $\theta$, at several times.}\label{Vtheta}
\vspace{-0.8cm}
\end{figure}
\vspace*{-0.5cm}
\section{Lateral spreading and global deceleration}\label{result}
\vspace*{-0.25cm}
During a long phase, the outflow propagates mainly spherically because of the relativistic initial radial speeds. However, the pressure contrast between the hot shell and the external medium causes a lateral spreading at the edge. In the meridional direction  this consists of (1) a shock propagating in the external medium and (2) a rarefaction wave propagating in the shell towards the axis. Since the BM shell has a radially stratified pressure that declines sharply behind the front shock, the lateral flow expands mainly from the region near the head of the shell. The spreading speed decreases quickly from just behind this region, inducing Kelvin-Helmholtz instability in the wings of the shell. This is fully consistent with the results by \cite{ZhangMacFadyen09}, and a zoom into these details is provided in Fig.~\ref{Figall}. This Kelvin-Helmholtz development does not affect shell dynamics, since the characteristic time for the lateral flow is longer than the timescale found for the radial deceleration.

The rarefaction wave that forms propagates inside the shell at the fluid's rest-frame sound speed. This wave extracts energy gradually, first from the edge and ultimately from the axial region. During the relativistic phase, the shell becomes structured meridionally, and energy, pressure, and density ultimately decrease from axis to edge. Deceleration occurs on the meridionally growing part of the shell affected by the wave, as quantified in Fig.~\ref{Figlfac}. This meridional stratification of the blast wave makes the edge of the shell decelerate faster than the inner part, bending the shell away from its spherical shape, as shown in Fig.~\ref{Blast_R}. 
This deformation becomes pronounced when the radial speed drops to about of the lateral speed, and increases the blast-wave working surface and the distance traversed by spreading matter. This is all in full agreement with previous simulations~\citep{Granotetal01,cannizzoetal04,ZhangMacFadyen09}. The rarefaction front reaches the axis after $t_{\rm local}\,\simeq\, 115.536\, {\rm days}$. By this time, the blast-wave has decelerated to $\gamma=1.09$ at the axis (near-Newtonian phase). 
In Fig.~\ref{Vtheta}, we quantify the meridional speed at the front of the shell for several times, to illustrate the propagation of the rarefaction wave. 
The meridional stratification of the blast wave decays when the Lorentz factor decreases to the point that causality links all parts of the shell together. 
In this near-Newtonian phase, energy becomes uniformly distributed throughout the shell.

\vspace{-0.6cm}
\section{Instability development in relativistic shocks}\label{result1}
\begin{figure}[h]
\begin{center}
{\resizebox{0.95\columnwidth}{4.3cm}{\includegraphics{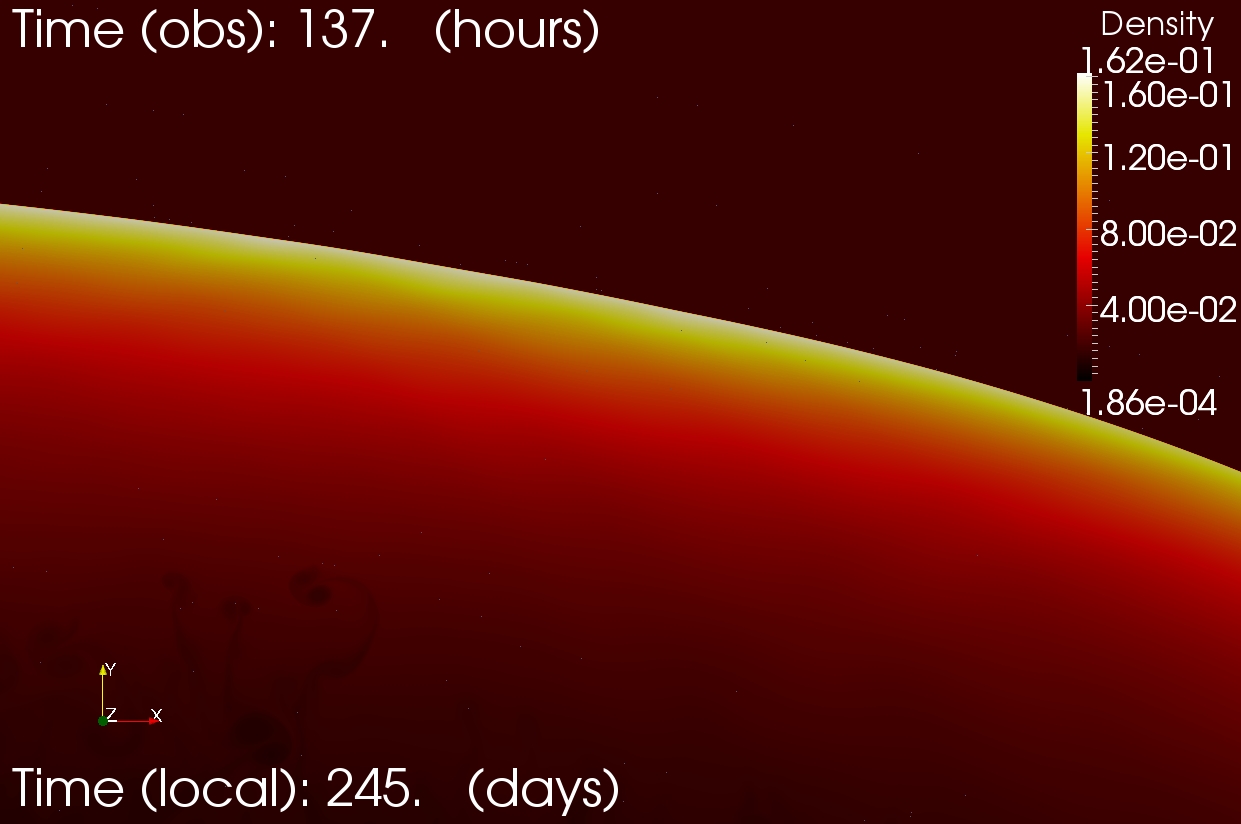}}}
{\resizebox{0.95\columnwidth}{4.3cm}{\includegraphics{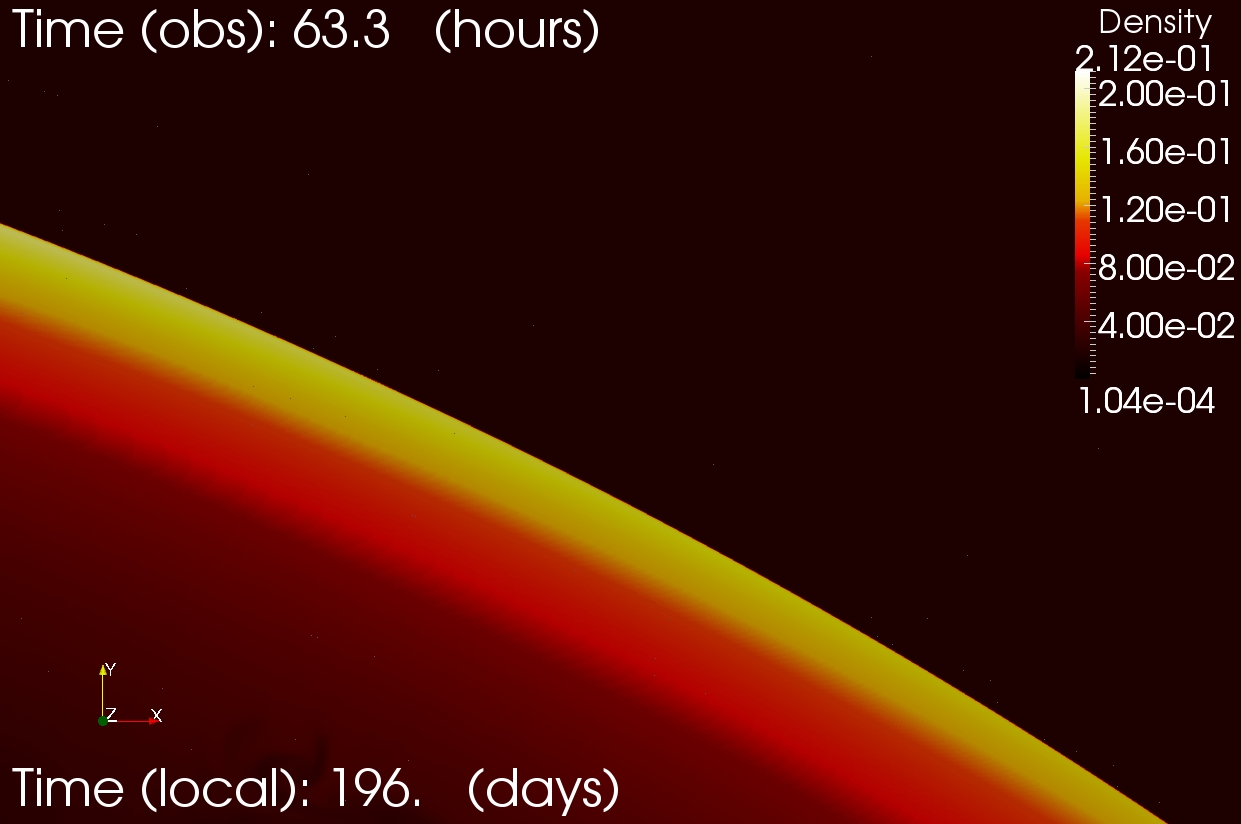}}}
{\resizebox{0.95\columnwidth}{4.3cm}{\includegraphics{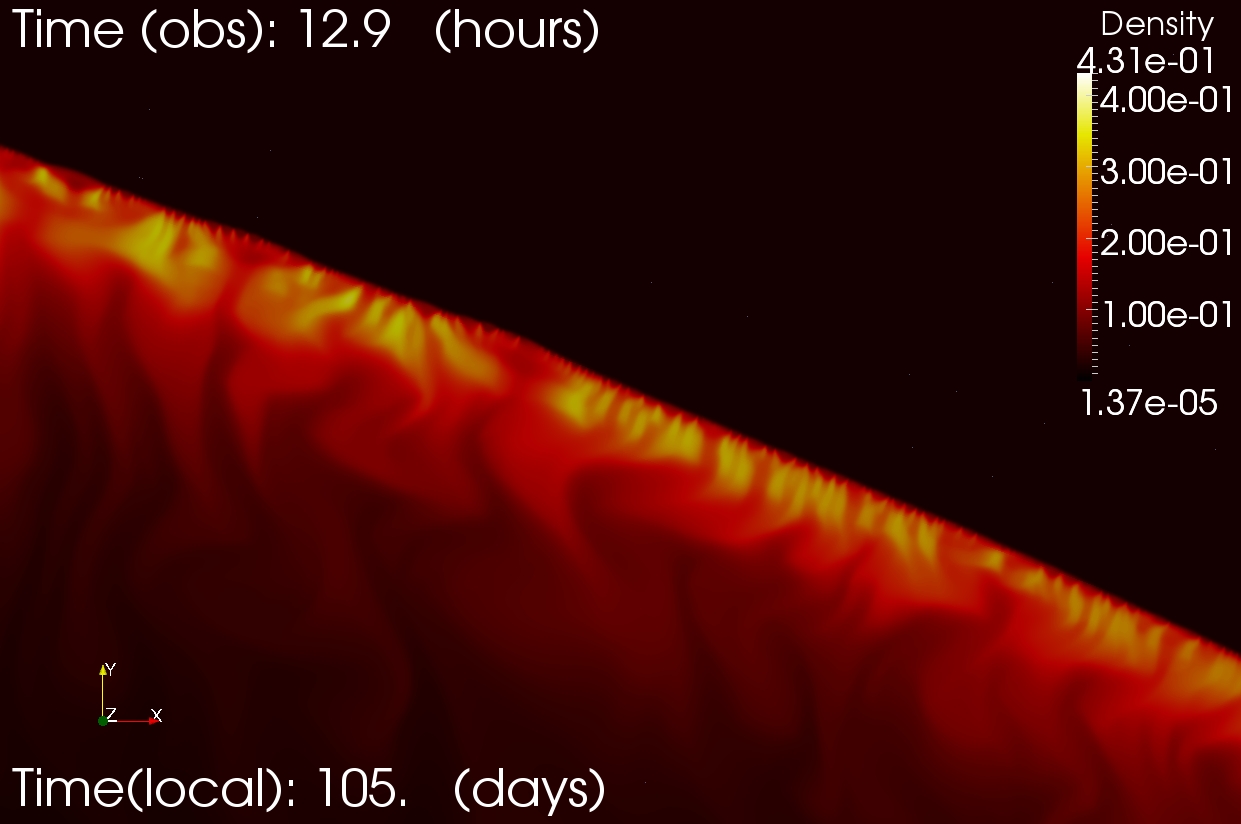}}}
{\resizebox{0.95\columnwidth}{4.3cm}{\includegraphics{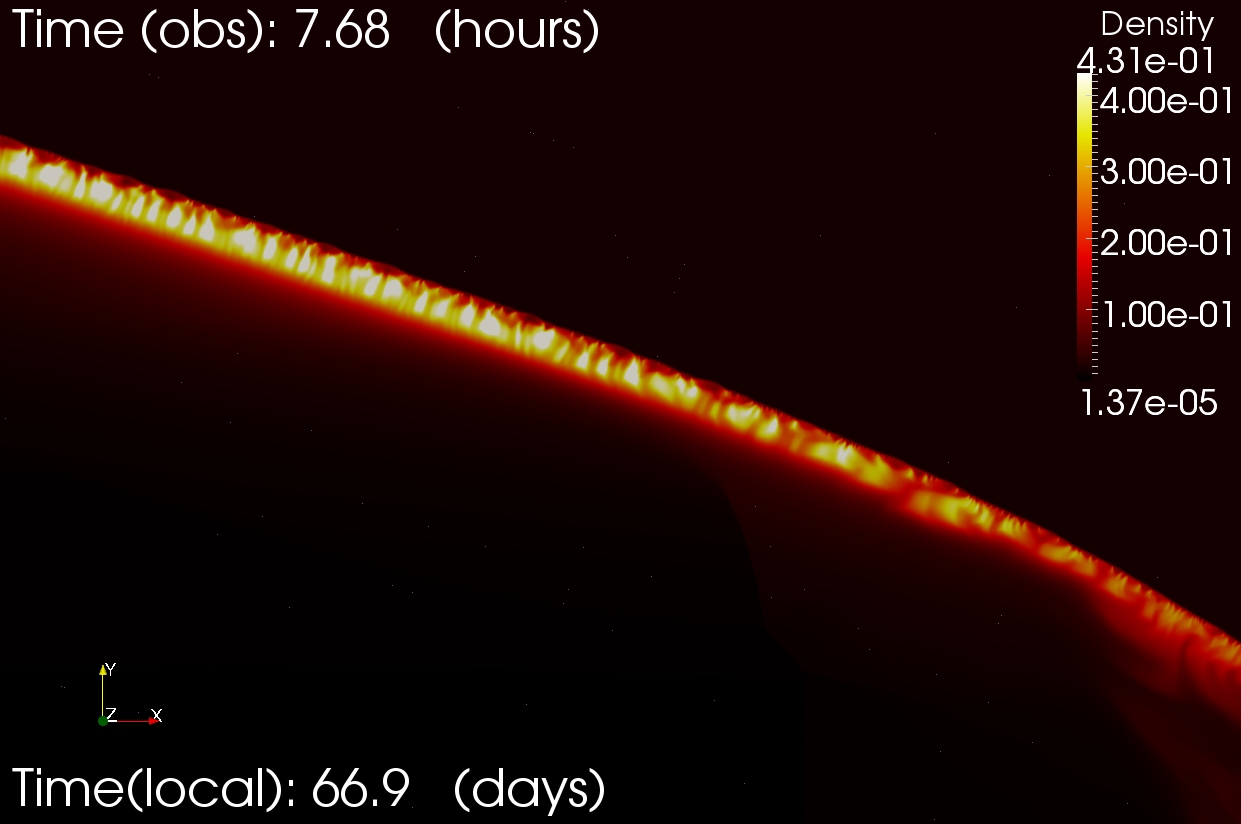}}}
{\resizebox{0.95\columnwidth}{4.3cm}{\includegraphics{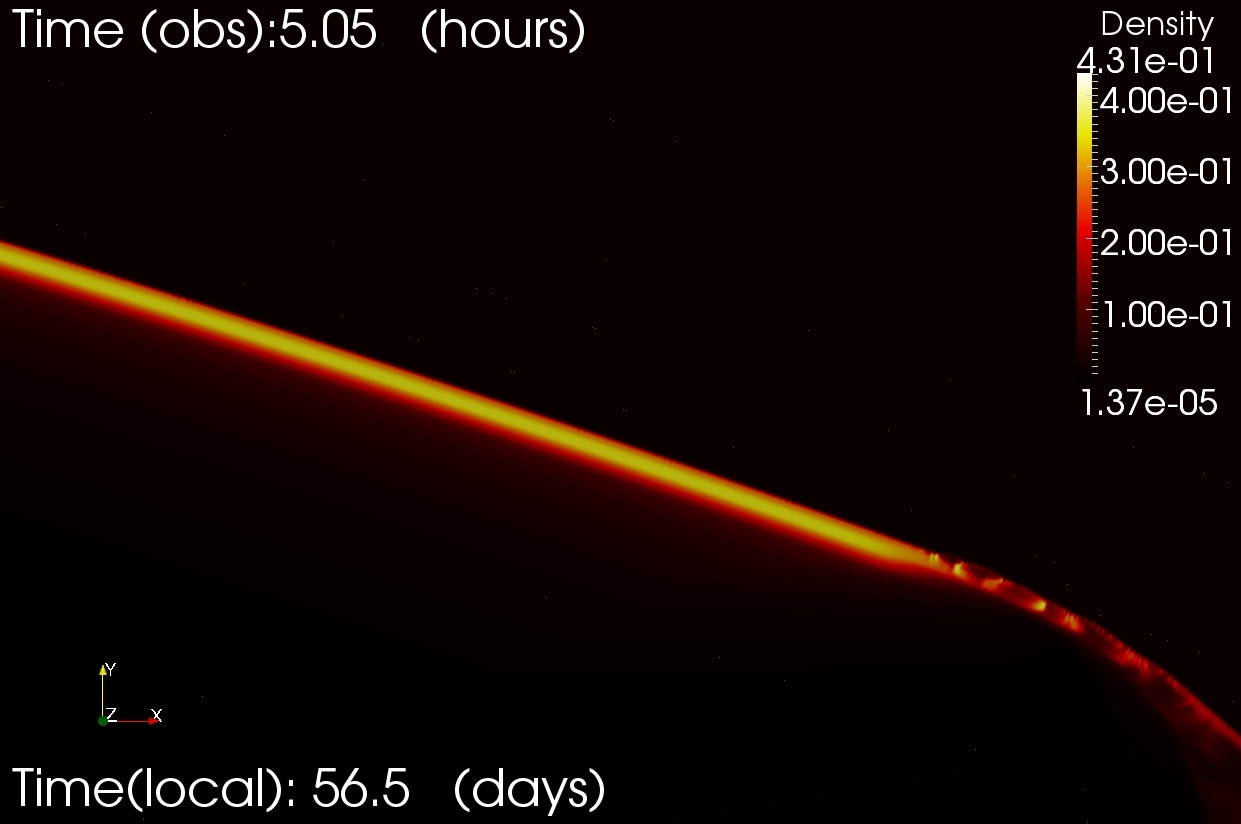}}}
\end{center}
\caption{The shell edge at various times, increasing from bottom to top. This shows the small-scale instability development and evolution.}\label{Fig1}
\end{figure}

\vspace{-0.25cm}
The most novel discoveries of our simulation are from the details of the initial evolutionary phase, where we now demonstrate that during the blast-wave deceleration, a ram pressure instability develops. This instability is found only while the blast wave speed remains relativistic, and disappears when its Lorentz factor becomes $\gamma< 3$ at time $t_{\rm local}\,\simeq\, 187 \,{\rm days}$ (or observer time $t_{\rm obs}\,\simeq \,53\, {\rm hours}$, as determined from $\Delta t_{\rm obs}=\Delta t_{\rm local}-\Delta R/c$). Figure~\ref{Fig1} shows its spontaneous appearance and disappearance, manifesting itself during some 100 days (local time). This instability induces small-scale bending of the blast front at wavelengths $\delta \theta \,\simeq\, 0.2^{\circ}$ (500 times larger than our cell meridional size, hence more than resolved properly). Detailed analysis indicates that each local front merely affects the minute region that is causally linked at this time, with causality angles around $\theta_{\rm causality} \simeq 7^\circ$ initially that increase during the evolution. The radial size of these ripples is given by $R\,\delta\theta \,\simeq\,  10^{-4}\, {\rm pc}$  at local time $67.5\,{\rm days}$. This is smaller than the thickness of the shell $\delta R \,\simeq \,3.7 \,\times\, 10^{-3}\, {\rm pc}$.
The instabilities develop as small-scale ripples at the shock front, and the higher thermal pressure inside them pushes the newly shocked external medium sideways to their left and right upstream. Fluid within the shell (behind the shock) is pushed downstream. This local structure generates vorticity in the post shock region, and is a prime candidate that can contribute to  magnetic field amplification. 

The instabilities depend, as in the classical case \citep{Vishniac83}, on internal pressure within the shell being thermal and isotropic, while external pressure is mainly ram pressure in the direction perpendicular to the shock front. This induces ripples, which grow in size as time proceeds, leading to shell fragmentation.
This fragmentation appears at a local time offset $\delta t_{\rm local}\approx 115$ hours after the start of our simulation, which is of the same order as the sound crossing time of the shell. At this time the shell has already decelerated to a Lorentz factor of order 10 and the local causality angle is reduced to about 7 degrees. 
This instability appears initially at the outer edge (see Fig.~\ref{Fig1} bottom), which is where the rarefaction wave first affects the shell.  
The edge thereby decelerates more quickly than the centre, is first to bend away from spherical shape and to reach the lower Lorentz factors. This is seen in the bottom panel of Fig.~\ref{Fig1}, while the second panel  upwards shows that later the entire shell develops local instabilities (even before the rarefaction wave has affected its evolution).
The size of the fragments increase because of the spherical expansion. Moreover, the density and Lorentz factor contrast between fragments and surroundings increase and more energy become concentrated in these the fragments. They appear as independent shells propagating downstream, until the Lorentz factor of the entire blast wave becomes smaller then $\gamma \leq 3$, or when they are reached by the inward propagating rarefaction wave. As the Lorentz factor of the shell gradually decreases, as well as the compression rate $\tau=\frac{\gamma \Gamma+1}{\Gamma-1}$, the sound crossing time for the shell increases. The shell ultimately returns to a configuration that is stable against these pressure-ram pressure instabilities.
In this near-Newtonian phase, the fragments diffuse and disappear, in a way that is consistent with linear analysis predictions \citep{Vishniac83}. During this phase, the blast wave propagates adiabatically with effective polytropic index of $\Gamma_{\rm eff}\;\simeq\; 5/3\,>\,1.3$. 
As the fragments are forced to move, they form tails in the rear of the BM shell. Our result is the first to show clearly that some of the as yet unexpected instability modes in BM solutions exist, decay slowly as the system evolves.
In this case, small-scale structure appears when the Lorentz factor is $10$, and  disappears when it reaches $3$.

As the rarefaction wave propagates towards the axis, it not only forces the fluid to spread towards the edge, but dampens these pressure-ram pressure instabilities. In the relativistic phase, the instabilities persist all the way to the axis. The rarefaction wave reaches the axis when the Lorentz factor is already low throughout, and the instabilities have  almost completely disappeared. 
After that, the shell follows the BM solution. However, since the rarefaction wave extracts energy, the energy density in the shell becomes lower than in the equivalent isotropic case and is not uniformly distributed. All these effects produce to a faster overall shell deceleration. At $t_{\rm local}=1166.27 \,{\rm days}$, along the axis the speed is $25\%$ lower than the value  obtained from the 1D analytical solution~\citep{BlandfordMcKee76}, and $31\%$ lower $20^\circ$ away from axis.

\vspace{-0.2cm}

\section{Conclusions}\label{discussion}
\vspace{-0.2cm}
We have carried out an extreme resolution simulation of propagating relativistic GRB jets in a constant density medium. We have concentrated  on dynamical aspects previously missed.
We have confirmed that the interface between the shell and the external medium develops a rarefaction wave propagating into the shell, which induces slow lateral spreading. The region affected by the rarefaction wave loses energy and decelerates more rapidly. From edge to axis this deceleration becomes gradually,  more efficient. The shell deforms, increasing the work surface and inducing an earlier shell deceleration.
Our main result concerns the development of a relativistic pressure-ram pressure instability at the front of the blast wave. This instability induces differences in the deceleration properties during the relativistic phase. It is a relativistic analogue of the instability discussed in Vishniac (1983), but in relativistic disguise where the compression rate is high even though the shock is adiabatic. The compression rate is then proportional to the Lorentz factor of the blast wave. Its ultimate fate is to decay slowly and disappear when the Lorentz factor decreases to $\gamma \leq 3$.  
Finally, it should be noted that because of the strong beaming of the nonthermal emission from the blast wave, the instability that forms at the shock front might be expected to have an effect on the shape of the observed light curve, since at the front shock this instability can induce instantaneous variation in Lorentz factor where $\gamma_{\rm max}\approx 3\gamma_{\rm min}$.

\begin{acknowledgements}
We acknowledge support from FWO, grant G.027708, and K.U.Leuven GOA/09/009.
Computations performed on JADE (CINES) in DARI project c2010046216 and on HPC VIC at K.U.Leuven. ZM acknowledges support from HPC Europa, project 228398.
Visualization used Paraview, {\tt www.paraview.org}. We thank the referee A. Ferrari for useful suggestions.
\end{acknowledgements}
\vspace{-0.8cm}
\bibliographystyle{aa}

\begin{thebibliography}{99}
\vspace{-0.2cm}

\bibitem[\protect\citeauthoryear{Blandford \& McKee}{1976}]{BlandfordMcKee76}Blandford, R.D., McKee, C. F., PhFl, 1976, 19, 1130
\bibitem[\protect\citeauthoryear{Bloom et al.}{2003}]{Bloometal03}Bloom, J.S., Frail, D.A., Kulkarni, S.R., 2003. \apj, 594, 674
\bibitem[\protect\citeauthoryear{Cannizzo et al.}{2004}]{cannizzoetal04}{Cannizzo}, J.K., {Gehrels}, N., {Vishniac}, E.T., 2004, \apj, 601, 380

\bibitem[\protect\citeauthoryear{Frail et al.}{1997}]{Frailetal97}Frail et al., 1997, Nature, 389, 261
\bibitem[\protect\citeauthoryear{Granot et al.}{2001}]{Granotetal01}Granot, J., Miller, M., Piran, T., Suen, W. M., Hughes, P.A., 2001, in Gamma Ray Bursts in the Afterglow Era, ed. E. Costa, F. Frontera, \& J. Hjorth
(Berlin: Springer), 312
\bibitem[\protect\citeauthoryear{Harten et al.}{1983}]{Hartenetal83}Harten, B., Lax, P.D., van Leer, B. 1983, SIAM Review, 25, 35
\bibitem[\protect\citeauthoryear{Kobayashi et al.}{1999}]{Kobayashietal99}Kobayashi, S., Piran, T., Sari, R., 1999, \apj, 513, 669
\bibitem[\protect\citeauthoryear{Kumar \& Granot}{2003}]{Kumar&Granot03}Kumar, P., Granot, J., 2003, \apj, 591, 1075
\bibitem[\protect\citeauthoryear{Lazzati et al.}{2004}]{Lazzatietal04}Lazzati, D., Rossi, E., Ghisellini, G., Rees, M.J., 2004, \mnras, 347, L1
\bibitem[\protect\citeauthoryear{Meliani et al.}{2004}]{Melianietal04}Meliani, Z., Sauty, C., Tsinganos, K., Vlahakis, N., 2004, \aap, 425, 773
\bibitem[\protect\citeauthoryear{Meliani \& Keppens}{2007}]{zaklet07}Meliani, Z., Keppens, R., 2007, \aap, 467, L41
\bibitem[\protect\citeauthoryear{Meliani et al.}{2007}]{Melianietal07}Meliani, Z.,
 Keppens, R., Casse, F., Giannios, D., 2007, \mnras, 376, 1189
\bibitem[\protect\citeauthoryear{M\'esz\'aros \& Rees}{1997}]{Meszaros&Rees97}M\'esz\'aros, P., Rees, M.J., 1997, \apj, 476, 232
\bibitem[\protect\citeauthoryear{M\"{u}ller et al.}{2010}]{Mulleretal10}M\"{u}ller, B.,  Janka, H.T., \&  Dimmelmeier, H., 2010, \apjl, 198, 104
\bibitem[\protect\citeauthoryear{Pandolfi \& D'Ambrosio}{2001}]{Pandolfi&DAmbrosio01}Pandolfi, M., D'Ambrosio, D., 2001, JCP, 166, 271 
\bibitem[\protect\citeauthoryear{Rhoads}{1999}]{Rhoads99}Rhoads, J.E., 1999, \apj, 525, 737
\bibitem[\protect\citeauthoryear{van Eerten et al.}{2010}]{eerten10}van Eerten, H.J., Meliani, Z., Wijers, R.A.M.J., Keppens, R., 2010, arXiv1005.3966
\bibitem[\protect\citeauthoryear{Vishniac}{1983}]{Vishniac83}Vishniac, E.T., 1983, \apj, 274, 152
\bibitem[\protect\citeauthoryear{Zhang \& MacFadyen}{2009}]{ZhangMacFadyen09}Zhang, W., MacFadyen, A.I., 2009, \apj, 698, 1261

\end{thebibliography}

\end{document}